# Can Formamide Be Formed on Interstellar Ice? An Atomistic Perspective

*Albert Rimola*[1*], *Dimitrios Skouteris*[2], *Nadia Balucani*[3,4,5], *Cecilia Ceccarelli*[5], *Joan Enrique-Romero*[5], *Vianney Taquet*[3], *Piero Ugliengo*[6*]

[1]Departament de Química, UniversitatAutònoma de Barcelona, 08193, Cataluña, Spain.

[2]Scuola Normale Superiore, Piazza dei Cavalieri 7, I-56126 Pisa, Italy

[3]INAF, OsservatorioAstrofisico di Arcetri, Largo E. Fermi 5, 50125 Firenze, Italy

[4]Dipartimento di Chimica, Biologia e Biotecnologie, Università di Perugia, ViaElce di Sotto 8, I-06123 Perugia, Italy

[5]Univ. Grenoble Alpes, CNRS, Institut de Planétologie et d'Astrophysique de Grenoble (IPAG), 38000 Grenoble, France

[6]Dipartimento di Chimica and Nanostructured Interfaces and Surfaces (NIS), UniversitàdegliStudi di Torino, Via P. Giuria 7, 10125 Torino, Italy.

**E-mail corresponding addresses:**

albert.rimola@uab.cat

piero.ugliengo@unito.it

**Abstract**

Interstellar formamide ($NH_2CHO$) has recently attracted significant attention due to its potential role as a molecular building block in the formation of precursor biomolecules relevant for the origin of life. Its formation, whether on the surfaces of the interstellar grains or in the gas phase, is currently debated. The present article presents new theoretical quantum chemical computations on possible $NH_2CHO$ formation routes in water-rich amorphous ices, simulated by a 33-$H_2O$-molecule cluster. We have considered three possible routes. The first one refers to a scenario used in several current astrochemical models, that is, the radical-radical association reaction between $NH_2$ and HCO. Our calculations show that formamide can indeed be formed, but in competition with formation of $NH_3$ and CO through a direct H transfer process. The final outcome of the $NH_2$ + HCO reactivity depends on the relative orientation of the two radicals on the ice surface. We then analyzed two other possibilities, suggested here for the first time: reaction of either HCN or CN with water molecules of the ice mantle. The reaction with HCN has been found to be characterized by large energy barriers and, therefore, cannot occur under the interstellar ice conditions. On the contrary, the reaction with the CN radical can occur, possibly leading through multiple steps to the formation of $NH_2CHO$. For this reaction, water molecules of the ice act as catalytic active sites since they help the H transfers involved in the process, thus reducing the energy barriers (compared to the gas-phase analogous reaction). Additionally, we apply a statistical model to estimate the reaction rate coefficient when considering the cluster of 33-$H_2O$-molecules as an isolated moiety with respect to the surrounding environment; i.e., the rest of the ice. Our conclusion is that CN quickly reacts with a molecule of amorphous ice and that it can synthesize formamide, even though the efficiency of the $NH_2CHO$ formation is difficult to estimate as it depends on the unknown number of ice water active sites and the fine details of energy transfer through the ice body itself. Our results have two important consequences on the modelling of interstellar surface-chemistry. First, the $H_2O$ molecules of the ice, usually considered as an inert support in astrochemical models, can instead react with active radicals, like CN, forming more complex species, and can also act as catalysts by helping H transfer processes. Second, most of the involved intermediate steps towards formamide formation on the 33-$H_2O$ molecule cluster are so fast that it is unlikely that the energy released in each of them can be dispersed in the entire ice body of the grain. In other words, the system cannot be fully equilibrated at the grain temperature in each intermediate step, as assumed in all current models, because the localized energy can promote endothermic or high barrier processes in small portions of the ice before complete equilibration. The timescale of energy redistribution within the ice molecules, a poorly characterized process, should be explicitly accounted for if a realistic model of grain surface chemistry is pursued.

## 1. Introduction

Formamide ($NH_2CHO$) was detected for the first time in the interstellar medium (ISM) in 1971 towards the massive star forming regions Sgr B2 and in Orion KL[1], but only recently its detection has attracted growing interest. Dedicated observational campaigns towards solar-type star forming regions have revealed the presence of formamide in a variety of pre-stellar and proto-stellar objects at different masses and evolutionary states.[2-7] In addition, formamide has also been detected in the comets Hale-Bopp[8] and Lovejoy.[9] The presence of formamide in these environments is important from the perspective of both astro- and prebiotic chemistry because it belongs to the family of interstellar complex organic molecules (iCOMs),[10] namely C-containing molecules with six or more atoms.[11] iCOMs are the next generation of molecules after simpler inorganic compounds (*e.g.*, $H_2O$, $NH_3$, $SO_2$) and they represent an important turning point in the sequence of chemical events that lead to the increase of molecular complexity in space. In particular, formamide is the simplest iCOM containing the four most important elements for biological systems (that is C, H, O, and N) and is the simplest compound holding a O=C-NH group, which is the same group joining amino acids into peptides. There is some experimental evidence[12-14] that formamide could have been the starting molecule in the primitive Earth for the synthesis of metabolic and genetic molecular building blocks, *i.e.*, amino acids, nitrogenous bases, acyclonucleosides, sugars, amino sugars, and carboxylic acids. Interestingly, in some of those synthetic routes, the presence of naturally-occurring minerals and metal oxides as catalysts is mandatory. Similar successful results were also obtained when the reactions occurred in the presence of meteoritic materials.[15]

There is currently a debate on how iCOMs are formed.[16-20] The two disputing paradigms predict formation either by reactions occurring in the gas phase or on surfaces of the icy dust grains. In the former scenario, hydrogenated species formed on the ice can partially desorb into the gas phase and subsequently react with other gaseous species to form iCOMs through a series of gas-phase processes.[17, 21] In the latter scenario, instead, homolytic dissociation of the frozen hydrogenated species produces radicals which, in turn, diffuse on the dust grain surfaces and finally recombine to form iCOMs.[22-23]

Formamide is not exempted from this debate. Astronomical observations show a good linear correlation between isocyanic acid (HNCO) and $NH_2CHO$, which might suggest hydrogenation of HNCO on ice as a plausible formation route leading to $NH_2CHO$:

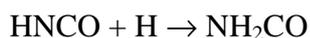

$$\text{HNCO} + \text{H} \rightarrow \text{NH}_2\text{CO}$$

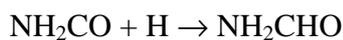

$$\text{NH}_2\text{CO} + \text{H} \rightarrow \text{NH}_2\text{CHO}$$

The first reaction was investigated as a bimolecular gas-phase process in accurate quantum chemical and statistical calculations by Nguyen et al.[24] This study indicated that the NH$_2$CO intermediate, rather than its possible isomers HNCOH and HNCHO, is mainly formed. Nevertheless, the NH$_2$CO intermediate quickly fragments into NH$_2$ + CO in the absence of a third body to stabilize it. Therefore, if we consider the same process occurring on the surface of interstellar icy mantles, in principle the NH$_2$CO intermediate could be stabilized by the interaction with the surrounding molecules and, subsequently, undergo further hydrogenation up to formamide formation. This suggestion has been disputed by recent experiments which have investigated hydrogenation of HNCO in its solid phase[25] and found no NH$_2$CHO to be formed in detectable amounts. This observation could be explained by considering that the H-abstraction reaction involving the NH$_2$CO radical (*i.e.*, NH$_2$CO + H → HNCO + H$_2$) prevails over further H-addition, thus leading the system back to HNCO (see Sec. 4.3.2). Furthermore, a recent theoretical work has indicated that the hydrogenation of HNCO on an amorphous solid water surface is a very slow process, slightly hampered rather than accelerated by the interaction with the surrounding molecules of the ice surface. Both the gas-phase and the amorphous solid water assisted reactions are rather inefficient with a rate coefficient of the order of $10^{-19}$ cm$^3$ s$^{-1}$ at a temperature of *ca*. 80 K.[26]

Gas-phase synthesis of formamide via the NH$_2$ + H$_2$CO → NH$_2$CHO + H reaction has recently gained support by quantum chemical and kinetic calculations.[27-29] Comparison of model predictions with astronomical observations showed that the proposed reaction can well reproduce the abundances of NH$_2$CHO in the cold envelope of the Sun-like protostar IRAS16293-2422 and the molecular shock L1157-B1, two very different interstellar objects[20]. More recent dedicated observations of the formamide spatial distribution towards the molecular shock L1157-B1 provided additional support to its formation from the $NH_2$ + $H_2CO$ gas-phase reaction.[20] Finally, the measured deuterated abundance ratio; *i.e.*, (2×NHDCHO + $NH_2CDO$)/$NH_2CHO$ towards the hot corino of IRAS16293-2422 are in perfect agreement with the predictions of the gas-phase chemistry.[29-31] However, an experimental investigation at low T is still desirable because of the presence of two transition states along the minimum energy path located very close in energy to the reactant asymptote.

For the sake of completeness, as an alternative to the above mentioned low energy scenarios, we mention here several experiments based on energetic processing of ice mantles which have resulted with the formation of $NH_2CHO$: i) frozen mixtures of $NH_3$ and CO exposed to 5 keV electrons;[32] ii) bombardment of ice mixtures of $H_2O:CH_4:N_2$, $H_2O:CH_4:NH_3$, and $CH_3OH:N_2$[33] with energetic ions; iii) simultaneous hydrogenation and UV-photolysis of CO, $H_2CO$ and $CH_3OH$-rich ices containing NO;[34] and iv) UV-photolysis and proton irradiation of mixtures of HCN with $H_2O$ and $H_2O:NH_3$ ices.[35]

In this article, we report a theoretical investigation of reactions on the grain surfaces which might, in principle, lead to formamide; i.e., the radical – radical association between $NH_2$ and HCO, and the reaction of the water of the ice with either HCN or CN, these two latter routes never considered so far. In order to verify whether they are efficient, we have performed dedicated quantum chemical calculations with the goal to evaluate their energetics and, provided they are feasible from that point of view, an estimate of their kinetics. The article is organized as follows. We first introduce the reactions considered in this work (§ 2) and the adopted methods (§ 3); we then provide the results of the calculations (§ 4) and discuss the astrophysical implications (§ 5). Conclusions are presented in § 6.

## 2. Proposed reactions to form formamide on the grain surfaces

In the large majority of current astrochemical models, formation of iCOMs is accounted for by the so-called warm-up mechanism.[36] According to it, several radicals, formed by the UV-photolysis of hydrogenated species on the grain surfaces, diffuse and react at temperatures (25-30 K), which are large enough to allow for some radical mobility.[22-23, 37] In this paradigm, formamide is alleged to be formed by the radical-radical association reaction:

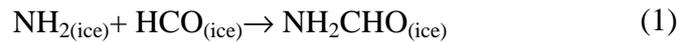

$$NH_{2(ice)} + HCO_{(ice)} \rightarrow NH_2CHO_{(ice)} \qquad (1)$$

The role of the ice surface is to dissipate the large energy amount liberated by the association reaction. However, after previous theoretical work by Enrique-Romero et al.,[18] it must be verified that reaction (1) actually leads to formamide rather than to other sets of products. In that work, the similar association reaction $CH_3 + HCO$ was shown to present formation of $CH_4 + CO$ as a competitive channel to $CH_3CHO$ formation. It is, thus, important to verify whether similar trends could also be operative in reaction (1). Therefore, characterization of reaction (1) occurring on a cluster of 33 $H_2O$ molecules is the first object of this work.

Searching for other reactions able to synthesize formamide on grain surfaces, we have considered here the possibility of reactions involving directly the water molecules of the icy mantles as *reactants*, as already invoked in different thermal reactions.[38] It is known, in fact, that isolated water molecules can react with radicals which are relatively abundant in the interstellar medium, including those normally considered in the model of Garrod *et al.*[22] As water already contains O and H, we have explored the possibility that water molecules of the ice react with N and C carriers. The two simplest and most abundant gaseous C- and N- bearing molecules in molecular clouds are HCN and CN: the average relative $HCN/H_2$ and $CN/H_2$ abundances in cold (~10 K) molecular clouds are about $2\times10^{-8}$ and $3\times10^{-10}$, respectively.[39] Given the low temperature, when HCN and CN encounter the icy mantle of dust grains, they remain frozen on the surface, but they could also undergo a chemical reaction. The reactions considered here are:

$$H_2O_{(ice)} + HCN_{(ice)} \rightarrow NH_2CHO_{(ice)} \tag{2}$$

$$H_2O_{(ice)} + CN_{(ice)} \rightarrow NH_2CO_{(ice)} \tag{3a}$$

$$NH_2CO_{(ice)} + H_{(ice)} \rightarrow NH_2CHO_{(ice)} \tag{3b}$$

$$NH_2CO_{(ice)} + H_2O_{(ice)} \rightarrow NH_2CHO_{(ice)} + OH_{(ice)} \tag{3c}$$

The focus of this work is on the actual chemical reactions; that is, processes in which chemical transformations take place, with the aim to provide an atomic-scale description and quantitative energetic data of these reactions. For this reason, other important processes involved in the overall process (namely, diffusion, photoreactivity or photodesorption) have not been accounted for.

## 3. Methods

*3.1 Electronic Structure Calculations*

Calculations were performed using the GAUSSIAN09 program package.[40] Stationary points were fully optimized (*i.e.*, both the ice cluster model and the reacting species) using the BHLYP hybrid density functional method.[41-42] Methods based on density functional theory (DFT) have been shown to be cost-effective and have been used to study a wide variety of closed-shell systems with great accuracy.[43-45] However, for open-shell systems, calculations carried out by some of us have demonstrated that functionals with a higher percentage of exact exchange, such as BHLYP, provide results in better agreement with wave-function based methods like the highly correlated CCSD(T).[46] This is because functionals based on the generalized gradient approach (GGA) or hybrid approaches with low percentages of exact exchange tend to overstabilize electron-delocalized situations as a result of the self-interaction error.[47] Despite this, to ensure the reliability of our results, we performed a preliminary calibration study comparing the energy barriers of the gas-phase reactions involving the species of interest calculated with BHLYP or other functionals and at the CCSD(T) level. Results indicate that indeed BHLYP energy barriers compare fairly well with the CCSD(T) ones (see Figure S1 and Table S1 of Supporting Information, SI). Geometry optimizations were performed with the standard 6-31+G(d,p) Pople basis set. The reaction energetics were refined by performing full BHLYP/6-311++G(d,p) single-point energy calculations on the BHLYP/6-31+G(d,p) optimized stationary points. Structures were characterized by the analytical calculation of the harmonic frequencies as minima (reactants and products) and saddle points (transition states). Zero-point energy (ZPE)-corrected values were obtained including thermochemical corrections computed at the BHLYP/6-31+G(d,p) level resulting from the standard rigid-rotor/harmonic-oscillator treatment[48] to the BHLYP/6-311++G(d,p) potential energy values. All the energetic values presented in this work include ZPE corrections. Binding energies of HCO, $NH_2$, HCN and CN with respect to their adsorbed state at the ice grain have been corrected for basis set superposition error (BSSE). Net atomic charges and electron spin densities on the atoms have been obtained by Mulliken and natural population analysis (NPA) of Weinhold *et al*.[49] Structures with doublet and triplet electronic states were simulated with open-shell calculations based on an unrestricted formalism; singlet biradical systems were calculated adopting an unrestricted broken-symmetry approach. An input example with the keywords used for this kind of calculations is provided in SI.

Reactions were studied on a 33-$H_2O$-cluster model representing the surface of an amorphous interstellar water ice. The structure and the procedure to obtain the cluster model are shown in Figure 1. From the crystalline structure of Ice XI (the proton-ordered low temperature analogue of the proton-disordered hexagonal ice, $I_h$), a minimal cluster model consisting of 18 water molecules was extracted (Figure 1A), which is the same cluster model employed by some of us to study glycine formation on water ice surface.[50] This model is relatively small to represent a large water ice surface, so we joined two of these minimal clusters (in which three water molecules were manually removed to facilitate the linkage of the clusters through hydrogen bonding), reaching the final cluster model of 33 water molecules (Figure 1B). This cluster model has also been used by some of us to simulate different reactions, such as successive hydrogenation of CO to form $CH_3OH$ or the reactivity between the HCO and $CH_3$ radicals, providing similar results with clusters of smaller size.[18, 51] Since this cluster was fully optimized in all the reactions studied, we checked the deformation of the cluster at each stationary point by performing single point energy calculations of the cluster without the reacting species and comparing this energy with the optimized bare cluster. The maximum deformation energy was found to be about 5 kJ mol$^{-1}$ (0.15 kJ mol$^{-1}$ per water molecule), and thus the deformation of the cluster does not exert any energetic influence in the calculated energy profiles.

*3.2 Kinetic calculations for the CN reaction*

By using the results of the electronic structure calculations, we have also provided an estimate of the rate coefficient for the most favorable scenario, that is, the one represented by the sequence of steps (3a)+(3c) (*vide infra*). To do so, we have assumed that the energy liberated in each exothermic step is instantaneously dissipated among all degrees of freedom of the CN + 33-$H_2O$ cluster, which is, however, considered as an adiabatic system which does not exchange energy with the surrounding environment; i.e., the rest of the icy mantle. This is an oversimplified vision, as more water molecules could participate in spreading the energy liberated by the first interaction of the CN radical with the water molecules. However, the further away an $H_2O$ molecule of the ice is from the reaction site, the less probable will be its involvement in the energy redistribution because of the competing fast reactions. In other words, we do not model the energy flowing from the exothermic reactive site to all water ice molecules, but rather treat it in an approximate way by considering it totally efficient for the closest 33 molecules and totally inefficient for the rest of farther water molecules.

We have adapted a kinetic code developed in-house by some of us to calculate rate coefficients of gas phase bimolecular reactions[52-54] and based on a Ramsperger-Rice-Kassel-Marcus (RRKM) scheme. Each of the steps involved in the adopted mechanism is treated as unimolecular, with the whole cluster being considered as a single "supermolecule", isolated from its surrounding environment and for which the intramolecular vibrational redistribution approximation holds. Three fundamental assumptions are made here: 1) the (3a) and (3c) steps proceed through a well-defined transition state, a "no-return" configuration which, when reached, is bound to lead to products; 2) the total energy of the cluster is constant and statistically distributed among all available degrees of freedom (300 when considering the 101 atoms of the cluster); after each step, the energy redistribution within the cluster occurs in a very short time but remains confined to the cluster itself; 3) the sequence ends with the desorption from the ice of one of the two products; i.e., either $NH_2CHO$ or OH, which as one can see below, OH being more likely because of its lower binding energy.

The degrees of freedom of the cluster comprise all the vibrations of the 33 water molecules (including intermolecular vibrations), vibration of the CN radical and overall rotations. The rate for each unimolecular step at a given total energy is given by the expression $(E) = \frac{N(E)}{h\rho(E)}$, where N(E) is the sum of states of the transition state (the number of states below the energy E), ρ(E) is the reactant density of states and $h$ is Planck's constant. The rate constants for all unimolecular steps are subsequently combined using the steady-state approximation to arrive at the overall rate constant for product formation and, finally, the rate constants are Boltzmann averaged to derive rate constants at specific temperatures.

## 4. Results

*4.1 The reaction $NH_2$+ HCO →$NH_2HCO$*

As illustrated in the Introduction, one of the proposed reactions to form $NH_2CHO$ on dust grain surfaces is the radical-radical association between $NH_2$ and HCO.[22] As for other radical-radical association reactions, the need to adopt the on-surface paradigm comes from the fact that these processes are largely exoergic. The dust grain acts as a third body that dissipates the energy released during the formation of the new bond, thus preventing the dissociation of the newly formed species back to reactants or towards other products. As already mentioned, however, a recent theoretical work carried out by some of us[18] indicated that the interaction of the two radical species on the amorphous water ice surface does not necessarily lead to the coupling of the unpaired electrons of the two radicals, but H transfer reactions can also occur.

Roughly speaking, surface reactions can take place through two possible mechanisms: i) Langmuir-Hinshelwood (LH), or Eley-Rideal (ER). In LH, the reacting species first adsorb on the grain surfaces, diffuse and, once they encounter one another, react, forming the final product. In ER, only one reactant is adsorbed on the surface, while the other comes from the gas phase to directly react with the adsorbed species. These two surface mechanisms represent the extreme regimes in terms of coverage; *i. e.*, ER can only work in high coverage regimes (the surface is plenty of adsorbed species), while LH is valid in a low coverage regime. The LH mechanism is usually considered the dominant process in the ISM because of its low atomic densities, leading to the formation of long-lasting adsorbates on the surface. Thus, we have considered the reaction of $NH_2$ and HCO through an LH mechanism, namely, when they are co-adsorbed on the 33-$H_2O$ ice cluster model.

To obtain an initial guess-structure for the reactivity between $NH_2$ and HCO, we have first performed calculations on the adsorption of the two radicals on the water ice cluster as separated species. The optimized geometries are shown in Figure 2A. It is worth mentioning that these structures represent one of the many possibilities in which the two radical species get adsorbed on the ice grain. However, they were built by maximizing the interactions, particularly the hydrogen bonds (H-bonds) that engage the radicals with the surface. In the resulting structures, $NH_2$ and HCO act as both donor and acceptor H-bond groups, thus forming two H-bonds each. The calculated binding energies for HCO and $NH_2$ on these specific sites are 17.5 and 33.5 kJ mol$^{-1}$ (equal to 2103 and 4026 K), which are in good agreement with the values reported by Wakelam *et al.*[55] Remarkably, for HCO adsorption, a similar structure was already calculated by some of us[18] on a smaller cluster model (18-water molecules) with a very similar binding energy (19.4kJ mol$^{-1}$). Sameera *et al.*[56] reported binding energies for HCO in different binding sites, ranging from 27 to 33 kJ mol$^{-1}$. These values were not corrected by ZPE corrections and, hence, this is probably the reason of the larger values compared to the value of our work.

In view of these adsorption states, one possible initial guess structure with the $NH_2$ and HCO radicals co-adsorbed is $NH_2$/HCO···W_1 of Figure 2B. Here, $NH_2$ and HCO adopt the same H-bonding patterns as in their isolated adsorption state. This structure was first calculated in a triplet state to prevent covalent bond formation and to obtain an optimized geometry based only on H-bonds between $NH_2$/HCO and $H_2O$ molecules of the ice surface. Subsequent geometry optimization in a singlet state (adopting an unrestricted broken-symmetry approach) yields the $NH_2$/HCO···W_1 biradical system, which has an absolute potential energy almost identical to the triplet-analogue state. A spin density analysis of the singlet biradical system indicates a spin distribution of about +1/-1 for $NH_2$/HCO; that is, the 100% spin density on $NH_2$ is up and the 100% of the spin density on HCO down. Table S2 of SI reports the spin densities of $NH_2$ and HCO in the triplet and biradical singlet states.

From $NH_2$/HCO···W_1, we have calculated the formation of $NH_2CHO$ by direct C-N coupling on the water ice surface. The calculated energy barrier for this process is very low (3 kJ mol$^{-1}$, see TS_$NH_2$HCO···W_1 of Figure 2B) and the final product is much more stable than the biradical reactant with a reaction energy of -381.5 kJ mol$^{-1}$ (see $NH_2$HCO···W_1). Thus, formation of formamide via $NH_2$/HCO coupling is indeed feasible under interstellar conditions, the only requisite being that the radicals have to encounter each other during their diffusion on the ice surface. It is also worth mentioning that, in view of the very low energy barrier, the $NH_2$/HCO···W_1 biradical is a steady-stable system; that is, it is stable but very reactive.

We have also considered the possibility that the $NH_2$ and HCO radicals, rather than forming $NH_2CHO$, can lead to the formation of $NH_3$ + CO. From the $NH_2/HCO \cdots W\_1$ structure, a direct H-transfer from HCO to $NH_2$ is not possible because both the H atom of HCO and the N atom of $NH_2$ are actually engaged by H-bond interactions with the ice surface. Thus, a possible path toward $NH_3$ + CO envisages the H atom of HCO to be transferred to $NH_2$ by the assistance of the ice water molecules through an H-relay mechanism. This process has been calculated and has a significantly larger energy barrier than the direct $NH_2CHO$ formation (about 35 kJ mol$^{-1}$, see structure TS_$NH_3/HCO \cdots W\_1$ of Figure 2B). In this transition state, 4 water molecules of the ice are involved in the H-transfer (all of them breaking/forming new bonds), a rather costly energy process.

Despite this, it is reasonable to think that formation of $NH_3$ + CO can take place via a direct H-transfer. We have calculated another reactant structure prone to drive this reaction: structure $NH_2/HCO \cdots W\_2$ of Figure 2C. As done for the biradical $NH_2/HCO \cdots W\_1$ system, $NH_2/HCO \cdots W\_2$ was first optimized in the triplet state and the resulting structure, subsequently optimized in the singlet state adopting an unrestricted broken-symmetry approach. However, the biradical $NH_2/HCO \cdots W\_2$ is not stable, since geometry optimization leads spontaneously to the formation of $NH_3$ + CO (see the sequence shown in Figure 2C). It is worth mentioning that the spin distribution of the initial wave function for $NH_2/HCO \cdots W\_2$ in its singlet state is +1/-1 for $NH_2/HCO$. Therefore, formation of $NH_3$ + CO through a direct H-transfer is a potential competitive reaction channel with formation of $NH_2CHO$ if the proper orientation of the radicals is achieved.

These results indicate that for both the formation of $NH_2CHO$ and of $NH_3$ + CO two requirements must be satisfied: i) $NH_2$ and HCO have to be in close proximity for the occurrence of the reactions, and ii) the orientation of the radicals is essential to drive one channel or the other. Therefore, assuming an almost random occupation of the surface sites, where the two initial radicals can be adsorbed in many different sites, and that the radicals diffuse on the surface, thus being in a non-equilibrium, dynamic regime, the reactivity of $NH_2$ and $HCO$ can lead to either $NH_2CHO$ or $NH_3 + CO$ formation. The diffusion of the radicals on the water ice surface will probably determine their relative orientation when they encounter, biasing the reactivity to one channel or the other.

*4.2 The reaction $H_2O + HCN \rightarrow NH_2CHO$*

The structure of an HCN molecule adsorbed on the water ice surface model is shown in Figure 3A. HCN is fully engaged by H-bond interactions with the surface, with a binding energy for this specific surface site of 46.8 kJ mol$^{-1}$(equal to 5269 K). This value is about 1.4 times larger than that derived by Wakelam *et al*.[55] We trace this difference back to the rather different cluster model adopted here to mimic the ice mantle compared to $H_2O$-HCN dimer with a single $H_2O$ molecule adopted by Wakelam *et al*. In our case, the adsorbed species experiences a more complex network of H-bond interactions compared to a single $H_2O$ molecule as done by Wakelam *et al*. Furthermore, it is well known that the strength of the H-bond is influenced by the cooperative effects within the ice particle due to the extended network of linked hydrogen bonded water molecules.[57] In essence, the terminal water molecules engaged in H-bonding with the HCN molecule (see Figure 3A) interact more strongly compared to their free state due to H-bond cooperativity within the cluster.

We have determined the variation of energy when HCN and one $H_2O$ molecule of the water cluster ice react to form $NH_2CHO$. The identified stationary points along the minimum energy path are represented in Figure 3B, while the energy content (including ZPE-corrections) is shown in Figure 4A, taking as the 0th-energy reference state AS1, which is the sum of the energies of HCN and the 33-$H_2O$ cluster model. The reaction takes place through two steps. The first step involves the nucleophilic attack of the O atom of the reacting $H_2O$ molecule to the C atom of HCN, with a simultaneous H transfer from the same $H_2O$ molecule to the N atom of HCN via the transition state TS_HNCHOH⋯W evolving toward the HNCHOH⋯W_1 intermediate. Remarkably, the H-transfer process is assisted by two additional $H_2O$ molecules of the ice. H-relay mechanisms of this kind are well-characterized catalytic processes, which reduce the energy barriers of H-transfer compared to the non-assisted ones.[58-60] Despite the water-assistance, the energy barrier of this step remains very high (about 167 kJ mol$^{-1}$), which in addition lies above in energy with respect to the initial AS1 state. The second step concerns the isomerization of HNCHOH into the actual $NH_2CHO$ by transferring an H-atom from the OH group to N. On the ice surface, this process is assisted by three water molecules; it exhibits an intrinsic energy barrier of about 29 kJ mol$^{-1}$. Although the reaction energies for the formation of the $NH_2CHO$ final species is favorable (with respect to the HCN⋯W reactant), the reaction is hindered by the high energy barrier of the first steps and no significant reactivity is expected between HCN and $H_2O$. It is worth mentioning that the present computational results are in agreement with recent experiments[61] in which the progressive warming of a $H_2O$:HCN ice mixture from 40 K to 180 K (to induce thermal activation) could not yield any reaction product. The ice components sublimate before reaction takes place.

*4.3 Energetics of reaction (3): $H_2O + CN \rightarrow NH_2CO \rightarrow NH_2CHO$*

CN is a radical species with an unpaired electron which can react with closed-shell molecules. We, therefore, expect a much more efficient reaction with the water molecules than the HCN case. In this line, experiments do indeed indicate an active low-temperature chemistry between hydroxyl radicals and water.[62]

We explored two main possibilities for the reaction scheme involving CN and the water molecules of ice, up to the formation of formamide. In the first one, we have considered reactions involving only CN and water molecules of the ice (reactions 3a+3c shown in § 2); in the second one, we have considered also the possibility that a free H atom, available on the ice surface, takes part into the global process (reactions 3a+3b shown § 2). In the conditions of the interstellar grains, free H atoms come from the gas phase. This is a situation similar to that assumed for the hydrogenation of several species, such as CO, on grain surfaces.[51, 63-64] H atoms land on the ice surface and scan it until they find a possible reactant. We will describe each of the two cases in the next sections.

*4.3.1 No free H atoms available*

The stationary points of the calculated reaction mechanism and its potential energy surface (including ZPE-corrections) are presented in Figure 5B and 4B, respectively. The 0$^{th}$ energy reference of the energy profile of Figure 4B is AS3, which is the sum of the energies of CN and the 33-H$_2$O-molecules cluster model. The CN···W structure of Figure 5A is the reactant state. This structure exhibits a hemi-bond between the CN species and a H$_2$O molecule of the ice with a distance of 1.876 Å. Hemi bonds are non-classical chemical bonds caused by the formation of a two-center three-electron bond. In this case the two centers are the O and the C atoms and the three electrons arise from one of the lone pairs of the $H_2O$ molecule and the unpaired electron of CN. The Mulliken spin densities on the O and the C atoms are +0.28 and +0.56, respectively (the remaining +0.16 is on the N atom) and the singly-occupied molecular orbital (SOMO) indicates that most of the unpaired electrondensity lies between the two centers (see Figure S2 of SI), which is the usual situation corresponding to a hemi-bonded structure. It is worth mentioning that the interaction of CN with one single water molecule also takes place through a hemi-bond interaction, which is 3.6 kJ mol$^{-1}$ more stable than the interaction through H-bond (see structures PREREACT and REACT of Figure S1 of SI). The calculated binding energy of CN for this specific water ice surface site is quite large, 76.6 kJ mol$^{-1}$ (equal to 8660 K). This value is more than 3 times larger than that derived in a recent work by Wakelam *et al.*[55] The origin of this discrepancy is probably given by the fact that the hemi-bond complex was not considered by Wakelam *et al*.

The global $CN_{(ice)}$ + $H_2O_{(ice)}$ reaction proceeds in three steps, as shown in Figure 5B and described here in detail.

Step 1: Similarly to the reaction between HCN and $H_2O$, the first step involves the complete bond formation between the O and the C atoms of the hemi-bonded structure, simultaneously to an H-transfer from the same reacting $H_2O$ water molecule to the N atom of CN (see TS_HNCOH⋯W) forming the HNCOH intermediate species (see HNCOH⋯W). The H transfer is assisted by two water molecules of the ice. The energetics of this step is very favorable, with a low energy barrier (16.1 kJ mol$^{-1}$) and the HNCOH⋯W structure being 81.2 kJ mol$^{-1}$ more stable than CN⋯W (see Figure 4B).

Step 2: The next step is the isomerization of the HNCOH intermediate yielding $NH_2CO$ (structure $NH_2CO$⋯W of Figure 5B, Step 2), in which the H atom of the OH group is transferred to the NH group. This isomerization is also assisted by three $H_2O$ molecules of the ice (structure TS_ $NH_2CO\cdots W$). The energetics of this step is even more favorable than the first one: the energy barrier is as low as 6.4 kJ mol$^{-1}$ and the reaction energy is -87.6 kJ mol$^{-1}$ (see Figure 4B).

Step 3: $NH_2CO$ is a radical-derivative of formamide, in which the H atom of the carbonyl (HCO) group is still missing. Thus, the third and final step (Figure 5B, Step 3) concerns the hydrogenation of the CO moiety to give $NH_2CHO$. In this case, one $H_2O$ molecule of the ice was adopted as the H atom source; *i.e.*, an H-transfer from a water molecule to $NH_2CO$ occurs (structure TS_$NH_2CHO\cdots W$), leaving an OH radical species at the water surface (see structure $NH_2CHO\cdots W/OH$). This direct H transfer is endothermic (+68.3 kJ mol$^{-1}$) and has a high-energy barrier of 111.6 kJ mol$^{-1}$ (see Figure 4B).

Remarkably, in the first two steps described above, water molecules act as catalysts since they actively participate to the H transfer process, dramatically reducing the energy barriers compared to the analogous gas-phase steps (see Figure S1 of SI). This reduction is due to the capability of water to help the H-transfer by simultaneously accepting and donating H atoms (H-relay mechanism). In this process, the role of the ice is to reduce the structural strain of the activated complex in the transition state compared to the gas-phase one. Clearly, the H-relay mechanism is only possible when the water molecules involved in the transfer are in a proper spatial orientation. Specifically, the H-bonding connections between the water molecules should allow the H atom to shuttle from the original position up to the "final destination" (see Figure 6A). When this is the case (as for Step 1 and Step 2), we label the responsible surface water molecules as "water active sites". When this mechanism is not operative; i.e., the H-bonding connections between the water molecules are truncated and do not allow the H-relays (see Figure 6B), all surface water molecules are considered as "water inactive sites". It is worth pointing out that determining the number of water active sites in a surface of interstellar ice is extremely difficult because of the large variety of possible ice structures in the amorphous state.

*4.3.2 One free H atom available*

We now consider the possibility that free H atoms are available on the ice surface so that they can hydrogenate the intermediates of the proposed CN + $H_2O$ reaction path. The first possible hydrogenation can occur after Step 1 of Figure 5B, when HNCOH is formed, while the second possible hydrogenation can occur after Step 2, when $NH_2CO$ is formed. We will refer to these two further steps as 4 and 5, respectively, even though they are not consecutive to the first three.

Step 4: Once HNCOH···W is formed, this structure can undergo H-addition with an incoming H atom, thus forming the HNCHOH species (see Figure 7A, Step 4a). This barrierless hydrogenation step is exothermic by -414.8 kJ mol$^{-1}$. It is followed by the isomerization of HNCHOH into the actual $NH_2CHO$ molecule through an H-transfer from the OH to the NH group (see Figure 7B, Step 4b). The isomerization was also computed adopting an H-transfer mechanism with the participation of three $H_2O$ molecules belonging to the ice water cluster. The isomerization energy is exothermic by -68.9 kJ mol$^{-1}$ but it has a non-negligible energy barrier of 40.8 kJ mol$^{-1}$ (see Figure 7B), significantly higher than the one associated with the isomerization of the HNCOH radical into $NH_2CO$ (step 2 of Figure 5B).

Step 5: To characterize the H addition to the $NH_2CO$, we have calculated the $NH_2CO\cdots W$ structure in the presence of one H atom at different positions on the surface to simulate different H adsorption sites. When geometry optimizations were carried out in a triplet electronic state to prevent covalent bond formation, H and $NH_2CO$ stay as co-adsorbates (see structures $H/NH_2CO\cdots W$ of Figure 8). In contrast, when we switched the electronic state to be singlet (within the unrestricted broken symmetry approach), the final resulting structure depends on the initial adsorption site of H. If the H atom is closer to the C atom than to the N atom, then $NH_2CHO$ is formed (see Figure 8A, Step 5a). In contrast, if the H atom is adsorbed closer to the N atom, a spontaneous (namely, during the optimization) H transfer from the NH group to the adsorbed H atom occurs, thus forming molecular $H_2$ and HNCO (see Figure 8B, Step 5b). This spontaneous hydrogen transfer between the $NH_2CO$ and H radicals hindering the formation of $NH_2CHO$ is in line with what was suggested in the above mentioned work by Noble *et al.*[25] related to the hydrogenation of solid HNCO. The calculated reaction energy for the $NH_2CO\cdots W + H \rightarrow NH_2CHO\cdots W$ process representing the H addition is -396.2 kJ mol$^{-1}$, whereas that for the $NH_2CO\cdots W + H \rightarrow HNCO\cdots W + H_2$ representing the H abstraction is -280.4 kJ mol$^{-1}$.

In summary, according to these calculations, in the presence of a free H atom, formamide can be formed through the Steps 1 → 2 → 5a and, much less efficiently, through Steps 1 → 2 → 4a → 4b because of the presence of a 40.8 kJ mol$^{-1}$ barrier in Step 4b.

*4.4 Kinetics of reaction (3): $H_2O + CN \rightarrow NH_2CO \rightarrow NH_2CHO$*

As shown in the previous section (§ 4.3), for the CN + $H_2O$ reaction, the most promising mechanism from an energetic point of view that can lead to the formation of formamide follows the sequence of steps 1 → 2 → 3. Because of that, we have estimated the rate coefficient for this pathway. The calculations were done by assuming that the energy released at each step is not lost by the system (where the system is formed by the initial CN radical plus the 33-$H_2O$-molecules of the cluster) for the entire duration of the process, but is efficiently scrambled among all the degrees of freedom of the cluster at each step. The reaction, indeed, starts with an initial energy budget of 76.6 kJ mol$^{-1}$, which is the energy suddenly liberated when CN first interacts with the water molecules of the surface. The total energy is conserved within the cluster throughout the whole transformation, but is redistributed among all degrees of freedom after every step. The energy produced by exothermic steps can be used by the system to overcome potential barriers or endothermic steps, as long as it remains within the 33-$H_2O$ cluster, as it happens in the case of isolated system. This approximation can be considered good if the sequence of reactions within the cluster is fast enough to prevent a significant loss of energy towards the surrounding environment during the duration of the process.

The rate constants as a function of temperature are shown in Figure 9. Within the temperature range of interest (10-50 K) the rate coefficient is *ca*. $1.9\times10^9$ s$^{-1}$, which implies an average reactive time of 0.5 ns. The lack of temperature dependence is expected because the system has an initial energy of -76.6 kJ mol$^{-1}$ (that is, the adsorption energy of CN on the water cluster). It can be noted that the average reactive time is quite smaller than the "typical" ones for gas-phase processes (in the ps and sub-ps range). Interestingly, back reaction is not expected to take place since the overall reaction is largely exoergic (including the desorption of the products, represented by AS4 and AS5 in Figure 4B). Particularly, the desorption barrier for OH is almost half to that of the back reaction (see AS5 of Figure 4B).

It is worth mentioning that this calculated rate coefficient is model-dependent, that is, if we apply the same formalism to smaller/larger clusters, the dissipation of the energy among the ice will be less/more efficient and accordingly overcoming energy barriers will be less/more difficult, thus increasing/decreasing the rate coefficient. The equidistribution of energy among a large number of degrees of freedom implies that the energy released as a result of an exothermic step is much less available to promote a subsequent step with an activation energy. More generally, the presence of a large number of degrees of freedom renders the energy difference of competing reactions a determining factor in the choice of the reaction path to follow, as the concentration of a given amount of energy in the reaction coordinate corresponds to a singularly high entropy of activation. This was especially seen considering the possible path of CN desorption from the cluster. The rate of CN desorption was calculated using a variational transition state theory (VTST) calculation based on various candidate transition states along the CN desorption path. From these calculations, the rate of CN desorption resulted to be around 9-10 orders of magnitude lower than that of Step 1 and hence, for all purposes, with no effect. In other words, once CN is captured by the icy surface, its most probable destiny is the reaction with water, promoted by the energy liberated during the adsorption step. In the assumptions of the present treatment, there are no other consuming pathways of CN radicals. If the cluster of water molecules loses a large part of its energy towards the surrounding environment in a timescale shorter than that associated to the reactions sequence, then our description is not valid anymore. On the other hand, it is to be noted that the assumption of a completely statistical distribution of energy among all degrees of freedom of the cluster could be compromised by the fact that such degrees of freedom are of different types, ranging from covalent bonds to hydrogen bonds and intermolecular interactions. As a result, equidistribution of energy might not be as efficient as hypothesized within the 33 molecules of the cluster and the liberated energy could be more localized at the reaction site making the global reaction much faster. This is commonly observed in rare-gas matrix experiments.[65] In this respect, the RRKM model represents a limiting case where energy diffusion is fast throughout the whole cluster and null outside it. Unfortunately, the energy redistribution after an exothermic reaction within the ice molecules is a poorly defined process, but its characterization is beyond the scope of this work. Recent attempts have been performed by considering the energy loss of $CH_4$, $CO_2$ and $H_2O$ with a 0.5-5 eV kinetic energy on the surface of crystalline and amorphous ice.[66-67] The focus of that work was, however, on the energy lost by those molecules interacting with the surface rather than on the dispersion of energy within the ice molecules.

## 5. Discussion

The calculations presented in this work indicate that formamide formation on the icy grain surfaces by the radical-radical association of $NH_2$ and HCO assumed in several current models is actually in competition with the formation of $CO+NH_3$. An additional pathway is started by the interaction/reaction between CN and some water molecules of the ice. The scheme of this possible chain of reactions is reported in Table 1 and shown graphically in Figure 10. In general, there are two possibilities:

  A. The chain of Steps 1 to 3 occurs without energy loss from the cluster, namely the reaction reaches the final Step 3 and forms $NH_2CHO$, because all the involved energy barriers can be overcome by the excess energy produced by the ongoing chemical steps in spite of the energy spreading along the 33 close-by molecules. Either OH or $NH_2CHO$ gets desorbed (more probably the former one due to its lower desorption energy, see AS4 and AS5 of Figure 4B).

B. At each step, the system has enough time to transfer the energy liberated by the reaction steps to the ice mantle. The energy available to continue the reaction sequence is only that associated with the temperature of the grains. Step 3 cannot occur because it is endothermic and the sequence actually stops at Step 2, with the formation of $NH_2CO$. Therefore, the only way toward formamide is by Steps 4 and 5.

Note that in the B case, the presence of free H atoms is mandatory to reach the final steps toward formamide. In the A case, assuming that CN is not consumed by other reactions on the grain surfaces, the amount of formamide will be the initial amount of CN on the grain surfaces scaled by the percentage of water active sites (defined in § 4.3.1). This is an upper bound to the quantity of formamide that can be synthesized on the grain surfaces, as in Steps 4 and 5 competing channels are also present (HNCHOH in Step 4 and HNCO in Step 5).

Let us analyze the two possibilities A and B assuming that all water molecules are active sites. In the A scenario, the computations reported in § 4.4 tell us that at 10 K the global rate coefficient is about $10^9$ $s^{-1}$. This has to be compared with the rate at which H atoms land on the grain surfaces. Assuming 10 K and a cloud with a H nuclei density of $10^4$ $cm^{-3}$, an average grain radius of 0.1 µm and the standard dust-to-gas mass ratio of 100, the accretion rate for atomic hydrogen per grain is about 0.1 $s^{-1}$.[68] Therefore, based on this simple argument, when a CN molecule lands on the grain surface it has a much larger probability to react with water, which is the major grain mantle species, than with an H atom. The same is true also for intermediate species. Note that even if just a fraction of water sites is active this conclusion still stands, because the water active sites intervene already in Step 1. Within the case B, the reaction proceeds rapidly up to the formation of $NH_2CO$ because of the low barriers involved; after that, it has to rely on the available H atoms.

In both cases, an important conclusion of our work is that the ice mantles of interstellar grains are not inert structureless surfaces: they are formed by water molecules that can easily react with active radicals when they get adsorbed on the surface and liberate the adsorption energy. These processes can be so fast (as in the case discussed here) that no other processes, like hydrogenation, can compete with them. To the best of our knowledge, this process has never been discussed in astrochemical models. Here, we have considered the case of the CN radicals, but analogous processes could also be efficient for other radicals.

Finally, our kinetic computations show that the CN-water reaction is fast enough, so that it is highly improbable that the system has time to transfer the energy liberated by the reaction from the 33-$H_2O$ cluster toward the rest of the ice mantle. In other words, large barriers with respect to the initial configuration, like the one of the Step 3, can be overcome because the system is not reset energetically at each step. This is also an important aspect that is usually not taken into account in current astrochemical models. An accurate model for energy distribution within the ice, explicitly considering the nature of the interaction among water molecules, is mandatory to establish the preeminence of scenario A or B.

## 6. Conclusions

In this work, different icy surface-assisted chemical paths leading to formamide formation have been studied by means of quantum chemical and kinetic calculations. A portion of the ice mantle has been modelled by a cluster system consisting of 33 $H_2O$ molecules. We verified here that the formation of formamide can proceed via the traditionally assumed association processes between the $NH_2$ and HCO radicals, but this reaction competes with the formation of $NH_3$ + CO through a direct H transfer from HCO to $NH_2$. Simulations indicate that the occurrence of one channel or the other depends on the relative orientation of the reacting radicals on the ice surface. However, two hitherto unexplored mechanisms featured by the reaction of either HCN or CN with water molecules belonging to the ice mantle have been studied. Our results indicate that the latter process can indeed lead to formamide formation or to that of its associated radical $NH_2CO$. For these favorable reactions, it is showed the paramount role of water, which: 1) acts as a reactant, since the first step concerns the C-O hemi-bond formation between CN radical and one water molecule of the ice mantle; 2) catalyzes steps 1 and 2 of the reaction (see Figure 10) through the hydrogen relay mechanism. In the absence of water, the structure of the transition states for these steps would be very strained, whereas water assistance allows a relevant reduction of strain and ultimately of the energy barriers; 3) is a possible H source to convert $NH_2CO$ into $NH_2CHO$. It is worth mentioning that, in order for this step to be operative, the nascent reaction energy of steps 1 and 2 should not be immediately dissipated throughout the ice mantle of the dust grain, but must be confined (at least partially) to a reduced number of nearby water molecules (33 in our model).

The above arguments and results have revealed that active radicals can indeed react directly with the water molecules of the ice, a process not usually taken into account in astrochemical models since water is considered to be an inert spectator. This might have profound implications on the fate of the radicals landing on the surfaces or also produced in the ice by dissociation because, very likely, these/some radicals will react with the water molecules, largely available in the ice substrate, before doing anything else. If this is confirmed on a large scale, the available astrochemical models might need to be substantially revised.

**7. Supporting Information**

Uncatalyzed gas-phase energy profile for the reaction of CN + $H_2O \rightarrow NH_2CO$ and the energetics provided by different electronic structure methods, SOMO image of the hemi-bond CN···W system, charges and spin densities, Gaussian09 input example for a biradical system calculation adopting an unrestricted broken symmetry approach, and optimized Cartesian coordinates and absolute energies of all the structures.

**8. Acknowledgments**

AR is indebted to "Ramón y Cajal" program, MINECO (project CTQ2017-89132-P) and DIUE (project 2017SGR1320). PU, NB and DS acknowledge the financial support by the Italian MIUR (Ministero dell'Istruzione, dell'Università e della Ricerca) and from Scuola Normale Superiore (project PRIN 2015, STARS in the CAOS - Simulation Tools for Astrochemical Reactivity and Spectroscopy in the Cyberinfrastructure for Astrochemical Organic Species, cod. 2015F59J3R). CC acknowledges funding from the European Research Council (ERC) under the European Union's Horizon 2020 research and innovation programme, for the Project "The Dawn of Organic Chemistry" (DOC), grant agreement No 741002. PU acknowledges $C^3S$ (http://c3s.unito.it) for generous allowance of computer time on the OCCAM computer. The use of the Catalonia Supercomputer Centre (CESCA) is gratefully acknowledged.

**Table 1.** Scheme of the reactions involved with the formation of formamide on the icy grain surfaces started by CN. The first half of the table reports the set of reactions where no free H atom participates, while in the second half free H atoms intervene in Steps 4 and 5. At each step, we report the reaction reactants and product and the associated energetic parameters. $E_{barr}$ is the activation energy barrier, $E_{reac}$ is the reaction energy (namely the difference of energy from the initial and the final state of each step), and $v^{\neq}$ is the transition frequency in cm$^{-1}$. All energies are in kJ mol$^{-1}$. Notes: *This is an isomerization process.

| | Step | Reactants | Products | $E_{barr}$ (kJ mol$^{-1}$) | $E_{reac}$ (kJ mol$^{-1}$) | $v^{\neq}$ (cm$^{-1}$) |
|---|---|---|---|---|---|---|
| No free H | 1 | CN + H$_2$O | HNCOH | 16.0 | -81.2 | 656.2 |
| | 2 | HNCOH* | NH$_2$CO | 6.4 | -87.6 | 449.5 |
| | 3 | NH$_2$CO + H$_2$O | NH$_2$CHO + OH | 111.6 | +68.3 | 1190.7 |
| Free H | 4a | HNCOH + H | HNCHOH | 0 | -414.8 | |
| | 4b | HNCHOH* | NH$_2$CHO | 40.8 | -68.9 | 858.3 |
| | 5a | NH$_2$CO + H | NH$_2$CHO | 0 | -396.2 | |
| | 5b | NH$_2$CO + H | HNCO + H$_2$ | 0 | -280.4 | |

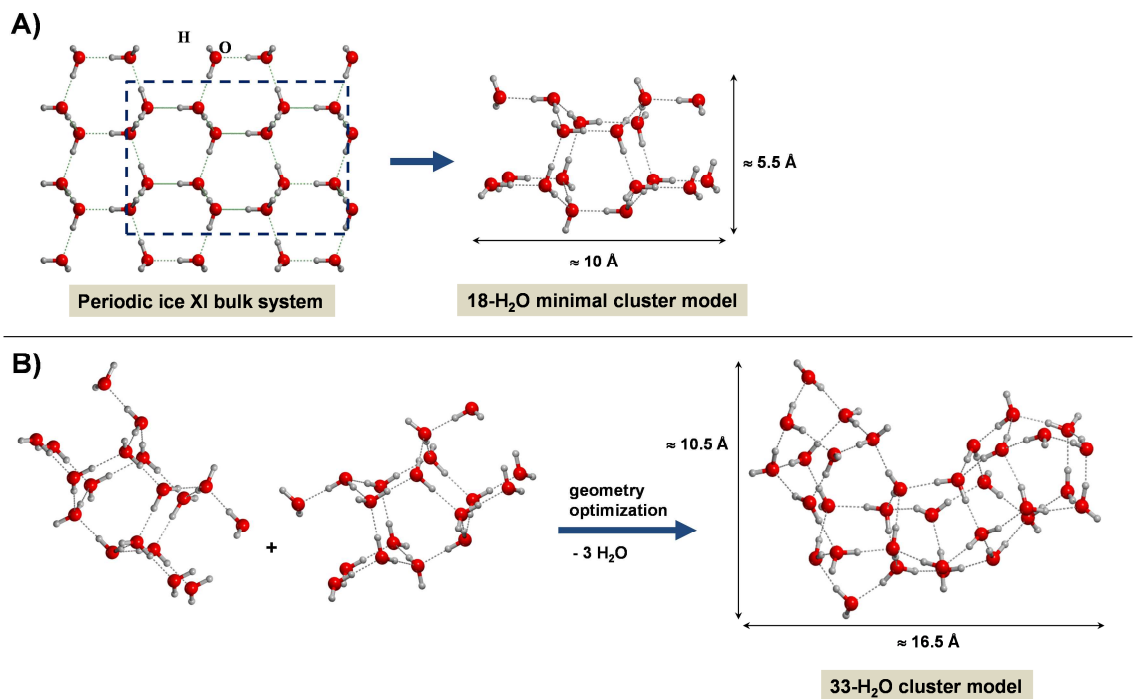

**Figure 1**.Procedure to generate the model cluster of the water ice surface. A) Extraction of the 18-H$_2$O minimal cluster from the periodic hexagonal ice bulk system. B) Linkage of the two minimal clusters to obtain the actual 32-H$_2$O cluster surface model.

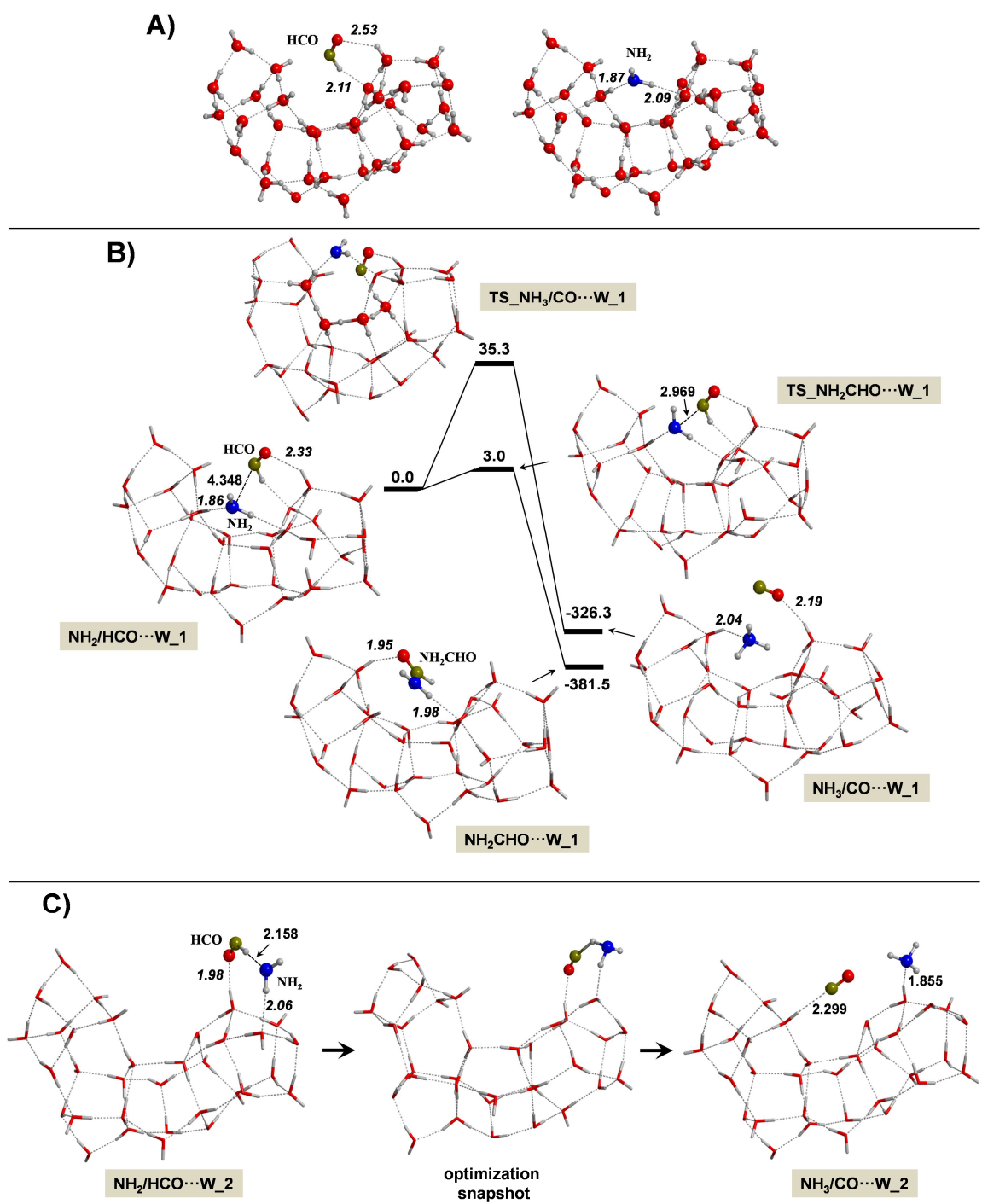

**Figure 2.** A) BHLYP/6-31+G(d,p) optimized geometries for the adsorption of the HCO (left) and $NH_2$ (right) radicals on the 33-$H_2O$ ice cluster model.. B) Potential energy surfaces (PESs) including zero-point energy (ZPE) corrections for the reactions of $NH_2CHO$ and $NH_3$ + CO formation on the water ice surface. Values are calculated by single-point energy calculations at BHLYP/6-311++G(d,p) onto the optimized BHLYP/6-31+G(d,p) geometries. ZPE corrections are those provided by the frequency calculations at BHLYP/6-31+G(d,p). Units are in kJ mol$^{-1}$. C) Snapshotsalong the geometry optimization at BHLYP/6-31+G(d,p) theory level involving a direct H transfer to form $NH_3$ and CO on the 33-$H_2O$ ice cluster model. Distances are in Å.

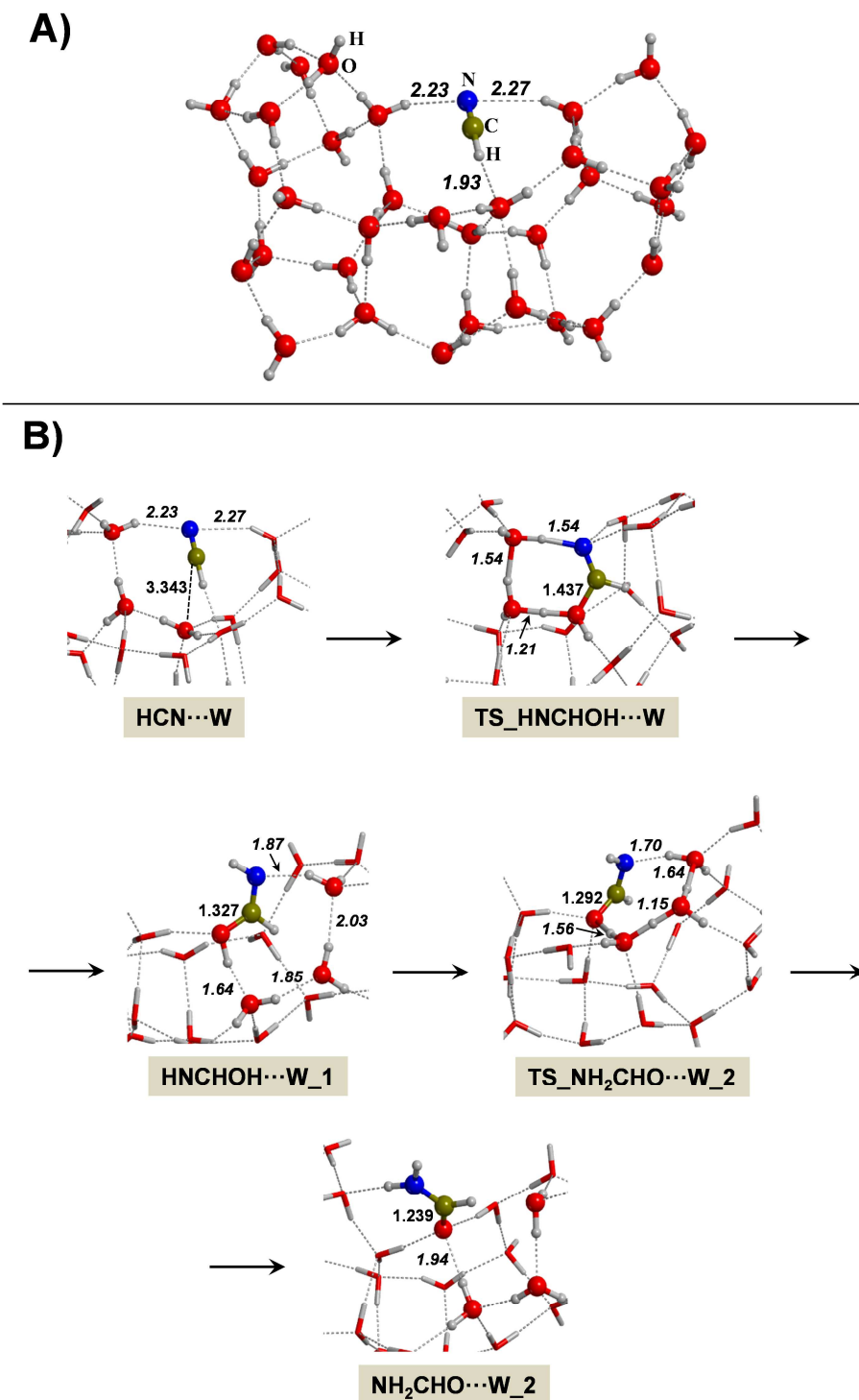

**Figure 3.** A) HCN adsorbed on the 33-$H_2O$ cluster ice surface model. B) BHLYP/6-31+G(d,p)-stationary points of the mechanism for the reaction of HCN with one $H_2O$ molecule of the ice. For the sake of clarity only the reactive part is shown. Bond distances are in Å.

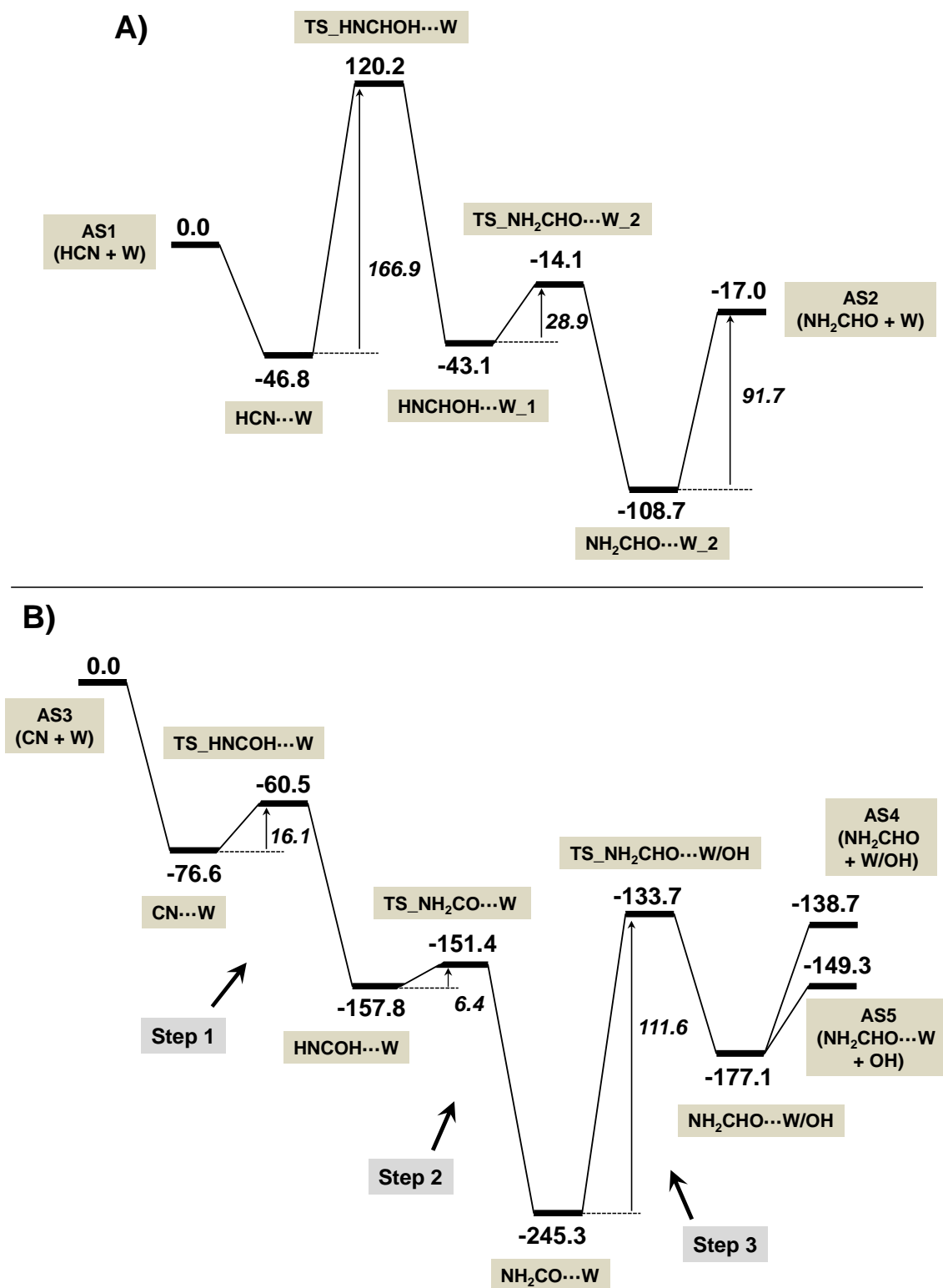

**Figure 4.** Potential energy surfaces (PESs) including zero-point energy (ZPE) corrections for the reactions of HCN (A) and CN (B) with one $H_2O$ molecule of the ice. Values are calculated by single-point energy calculations at BHLYP/6-311++G(d,p) onto the optimized BHLYP/6-31+G(d,p) geometries. ZPE corrections are those provided by the frequency calculations at BHLYP/6-31+G(d,p). Units are in kJ mol$^{-1}$.

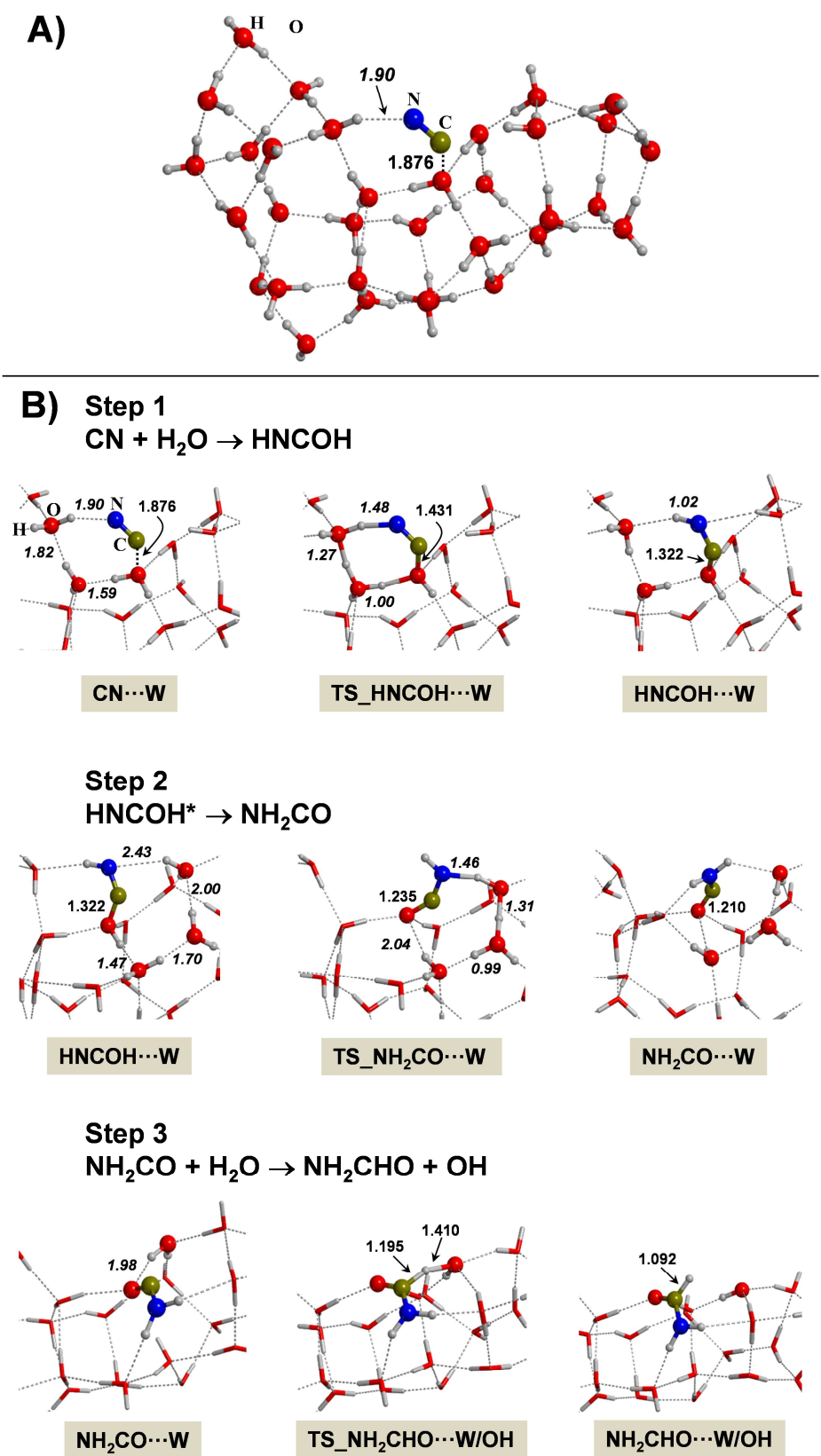

**Figure 5.** A) CN adsorbed on the 33-$H_2O$ cluster ice surface model. B) BHLYP/6-31+G(d,p)-stationary points of the mechanism for the reaction of CN with one $H_2O$ molecule of the ice. For the sake of clarity only the reactive part is shown. Bond distances are in Å.

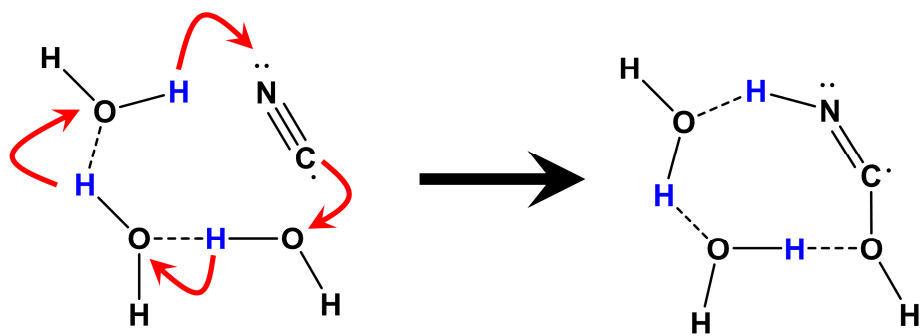

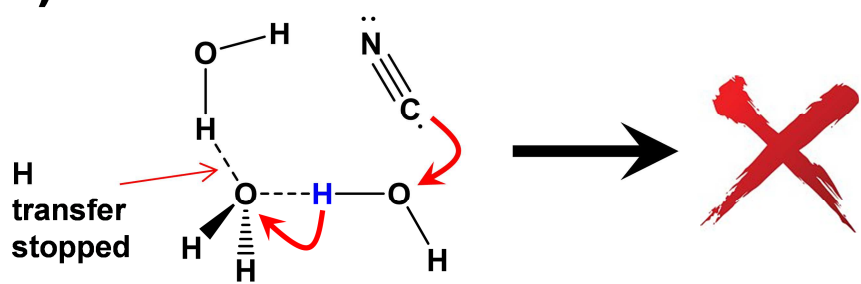

**Figure 6.** A) The H relay mechanism is operative. The three water molecules are considered "active sites". B) The H transfer is inhibited.

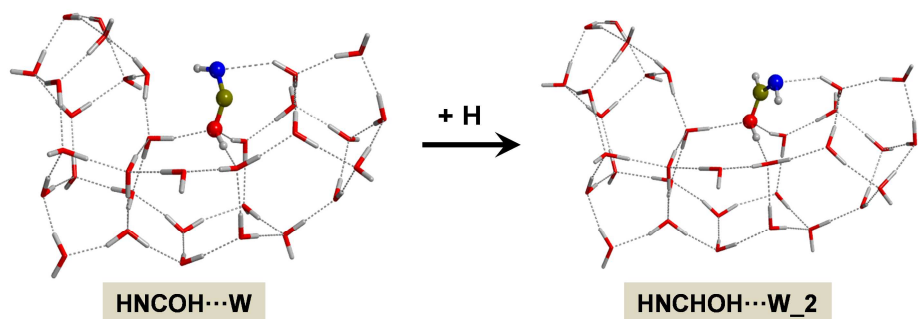

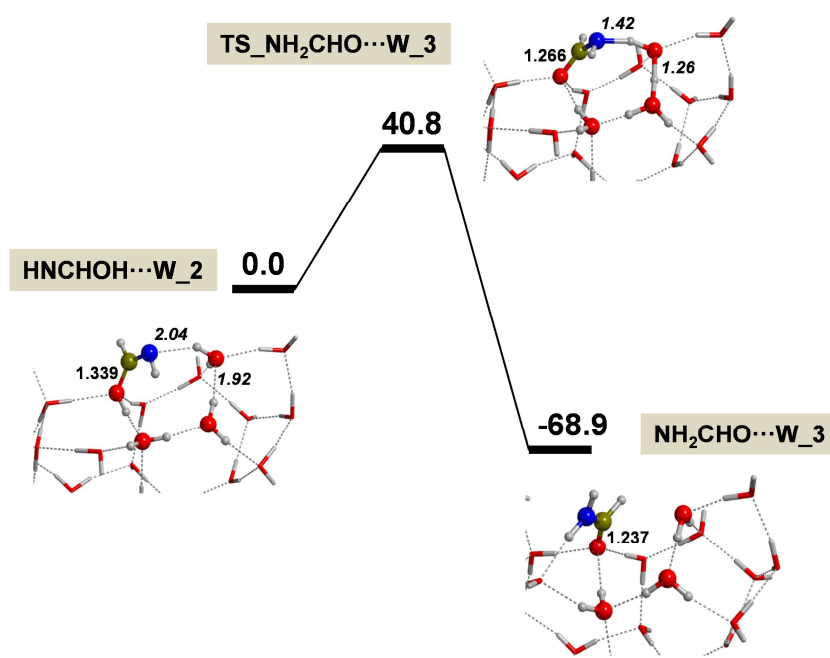

**Figure 7.** A) Reaction of the spontaneous H addition on the HNCOH⋯W structure. B) BHLYP/6-31+G(d,p)-stationary points of the HNCHOH⋯W_2 → NH$_2$CHO⋯W_2 isomerization on the water ice cluster model. For the sake of clarity only the reactive part is shown. Bond distances are in Å. C) Potential energy surface including zero-point energy (ZPE) corrections for the reaction shown in section B. Values are calculated by single-point energy calculations at BHLYP/6-311++G(d,p) onto the optimized BHLYP/6-31+G(d,p) geometries. ZPE corrections are those provided by the frequency calculations at BHLYP/6-31+G(d,p). Units are in kJ mol$^{-1}$.

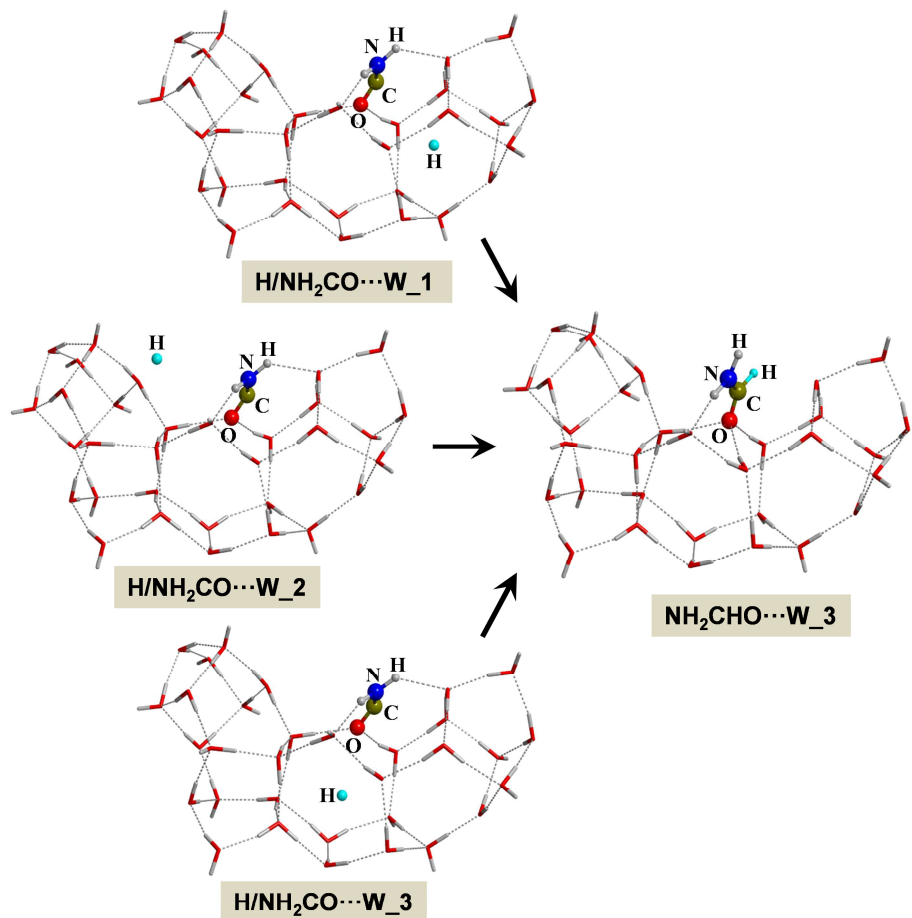

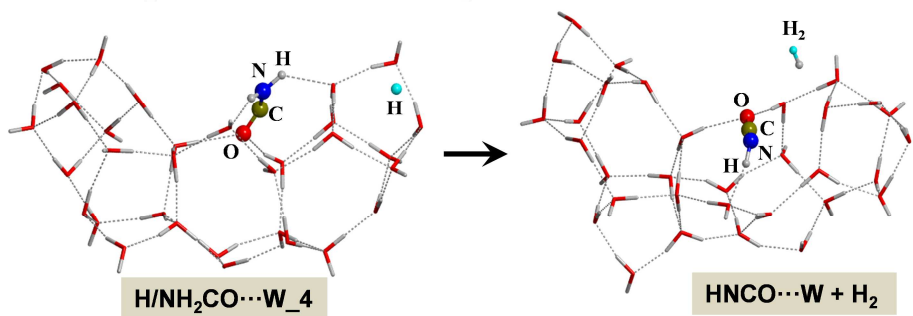

**Figure 8.** BHLYP/6-31+G(d,p)-optimized geometries of an H atom and $NH_2CO$ co-adsorbed on the 32-$H_2O$ surface cluster model in the triplet electronic state (structures on the left) and upon geometry relaxation in the singlet electronic state (structures at right). A) The resulting structure is formation of $NH_2CHO$ due to a spontaneous H coupling to the C atom during the optimization process. B) The resulting structure is formation of $HNCO + H_2$ due to a spontaneous H transfer from $NH_2CO$ to the H atom during the optimization process.

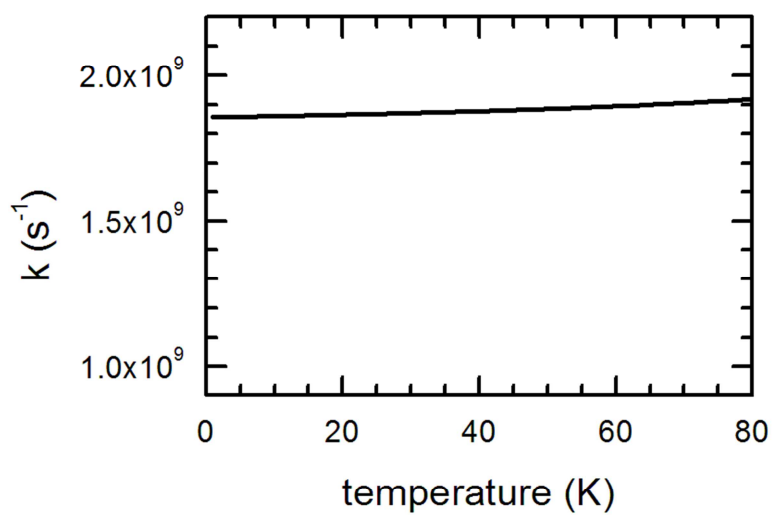

**Figure 9.** Unimolecular rate constant for the formation of formamide as a function of temperature.

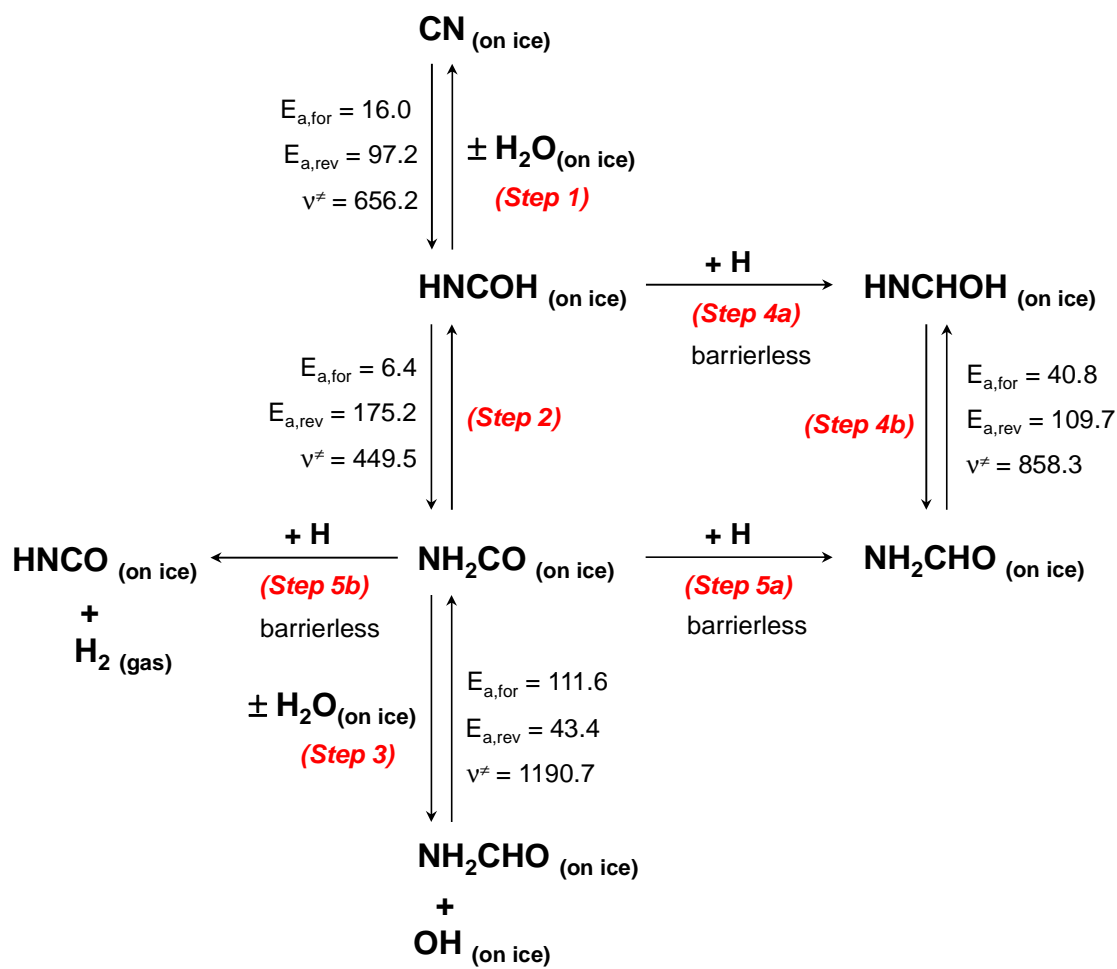

**Figure 10.** Scheme of the chain of reactions proposed to form $NH_2CHO$ from the reaction and interaction of CN with the water molecules of the ice. The energy barriers for the forward and the reverse reactions ($E_{a,for}$ and $E_{a,rev}$, in kJ mol$^{-1}$) alongside the transition frequencies ($\nu^{\neq}$, in cm$^{-1}$) obtained by the quantum chemical calculations are also shown. In this figure, it is assumed that energy is dispersed at each reaction step, so that the energy is computed as the difference of the initial and final (intermediate) single reaction step.

**For TOC Only**

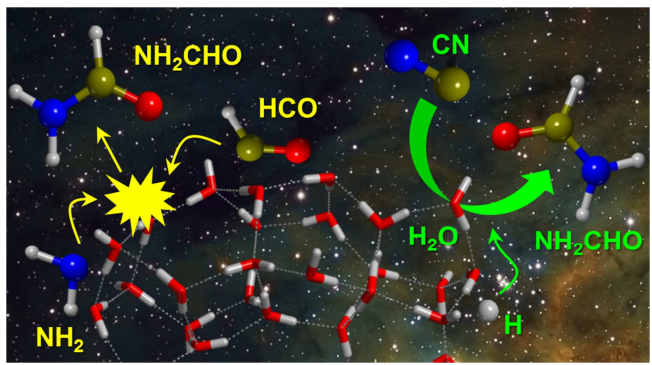